\documentclass[10pt]{article}

\usepackage[margin=0.75in]{geometry}
\usepackage{multicol} 
\usepackage{soul}
\usepackage[utf8]  {inputenc}
\usepackage{graphicx}
\usepackage{footnote}

\usepackage{empheq}
\usepackage{mathtools}
\usepackage{comment}

\usepackage{sidecap}
\sidecaptionvpos{figure}{t}  
\usepackage{wrapfig} 
\usepackage{amsmath}
\begin{document}

\begin{center}\large{\textbf{STUDYING EXOPLANETS IN THE RADIO FROM THE MOON}} \vspace{-0.2em}\end{center}
\begin{center}Jake D. Turner$^{1*}$, J.O. Burns$^{2}$, D. Rapetti$^{2,3}$, P. Zarka$^{4}$, J-M. Grießmeier$^{5}$, J. Bowman$^{6}$, G. Hallinan$^{7}$, J.	Hibbard$^{2}$, J. Jones$^{2}$, L. Lamy$^{4}$, C.K. Louis$^{4}$, R. Lovelace$^{1}$, N. Mahesh$^{7}$, R. Polidan$^{8}$, X. Zhang$^{4}$\\
$^{1}$Cornell University, $^{2}$CU Boulder, $^{3}$NASA Ames/USRA, $^{4}$Observatoire de Paris, $^{5}$Université d’Orléans/CNRS, $^{6}$ASU, $^{7}$Caltech, $^{8}$Lunar Resources, Inc., $^{*}$jaketurner@cornell.edu \\
{\scriptsize \noindent Submitted to the ``Key Non-Polar Destinations Across the Moon to Address Decadal-level Science Objectives with Human Explorers'' input call} 
\end{center}

%

\vspace{-0.5em}
\noindent\textbf{Introduction}: \\
Exoplanets with and without a magnetic field are predicted to form, behave, and evolve very differently$^{3}$.  Therefore, there is great need to directly constrain these fields to holistically understand the properties of exoplanets. Specifically, the observation of exoplanetary magnetic fields can yield valuable information regarding their interior structures, atmospheric dynamics, atmospheric escape, formation histories, and potential habitability$^{3,6,7,9,15,16,28,29}$. Thus, studying such fields aligns with the \textit{Worlds and Suns in Context} science theme in the Astro2020 Decadal Survey. Despite decades of searching, there is still no conclusive detection of an exoplanet's magnetic field$^{7}$. Although promising hints are starting to emerge$^{23,30}$.   

\begin{wrapfigure}{R}{0.5\textwidth}
  \begin{center}
   \vspace{-30pt}
   \includegraphics[width=0.5\textwidth]{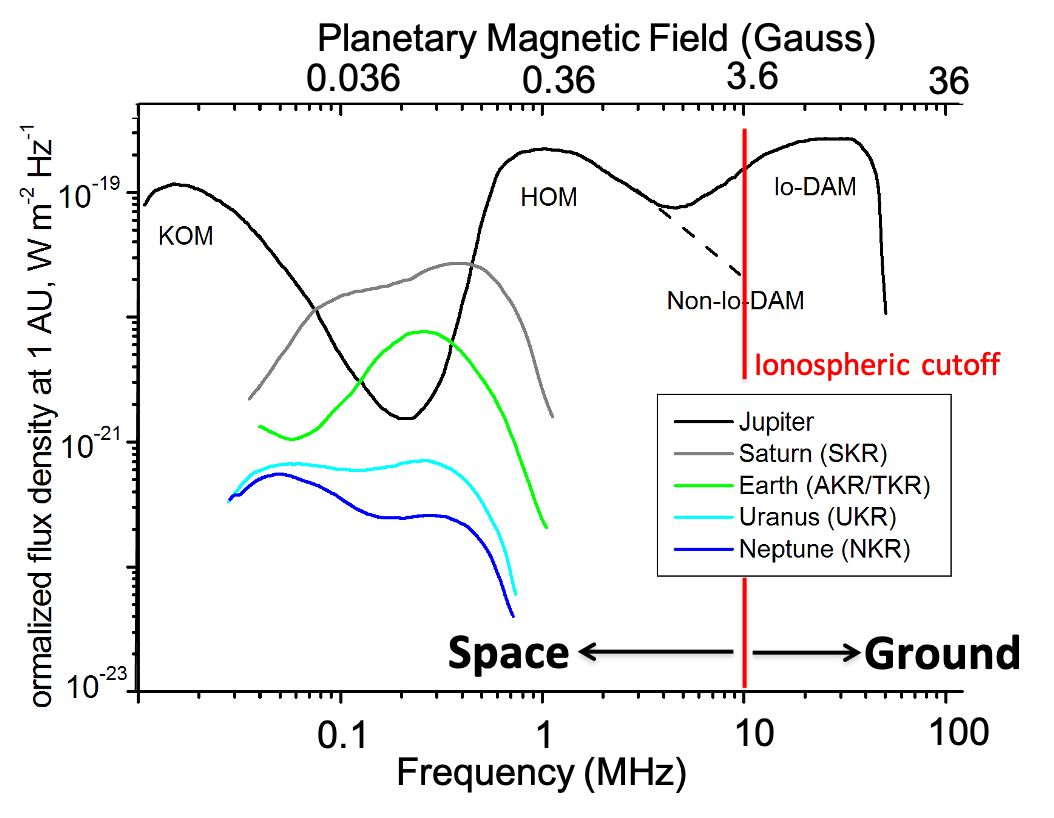} 
      \vspace{-30pt}
  \caption{Comparative radio spectra for the Solar System planets$^{26}$. This figure is adapted from [26].  \vspace{-2em}}
   \label{fig:Radio_SS}
  \end{center} 
\end{wrapfigure}

Observing planetary auroral radio emission is one of the most promising methods to detect exoplanetary magnetic fields$^{3,7}$. An exoplanet's magnetic field can be detected through radio emission from the planet generated by the electron-cyclotron maser instability (CMI). All the magnetized planets and moons in our Solar System emit or induce radio emissions using the CMI mechanism$^{26}$. CMI emission is highly circularly polarized, beamed, and time-variable$^{26}$ and occurs at the cyclotron frequency in the source region, $\nu$$_{g}$[MHz] = 2.8 $\times$ B$_{p}$[G] up to a maximum frequency that corresponds to the strength of the magnetic field near the planetary surface (B$_{p}$). In our Solar System, only Jupiter has a strong enough magnetic field to be studied from the ground (see Figure \ref{fig:Radio_SS}). Exoplanetary CMI radio emission is caused by electrons from the stellar wind interacting with the planetary magnetosphere$^{7,8,29}$. Additionally, radio observations will place informative constraints on the exoplanet's magnetospheric structure, space environment, orbital-inclination, rotation period, and any presence of extrasolar moons$^{13,29}$. All of these parameters are extremely difficult to study otherwise.

\begin{SCfigure}
\vspace{-1em}
\centering \includegraphics[width=0.5\textwidth]{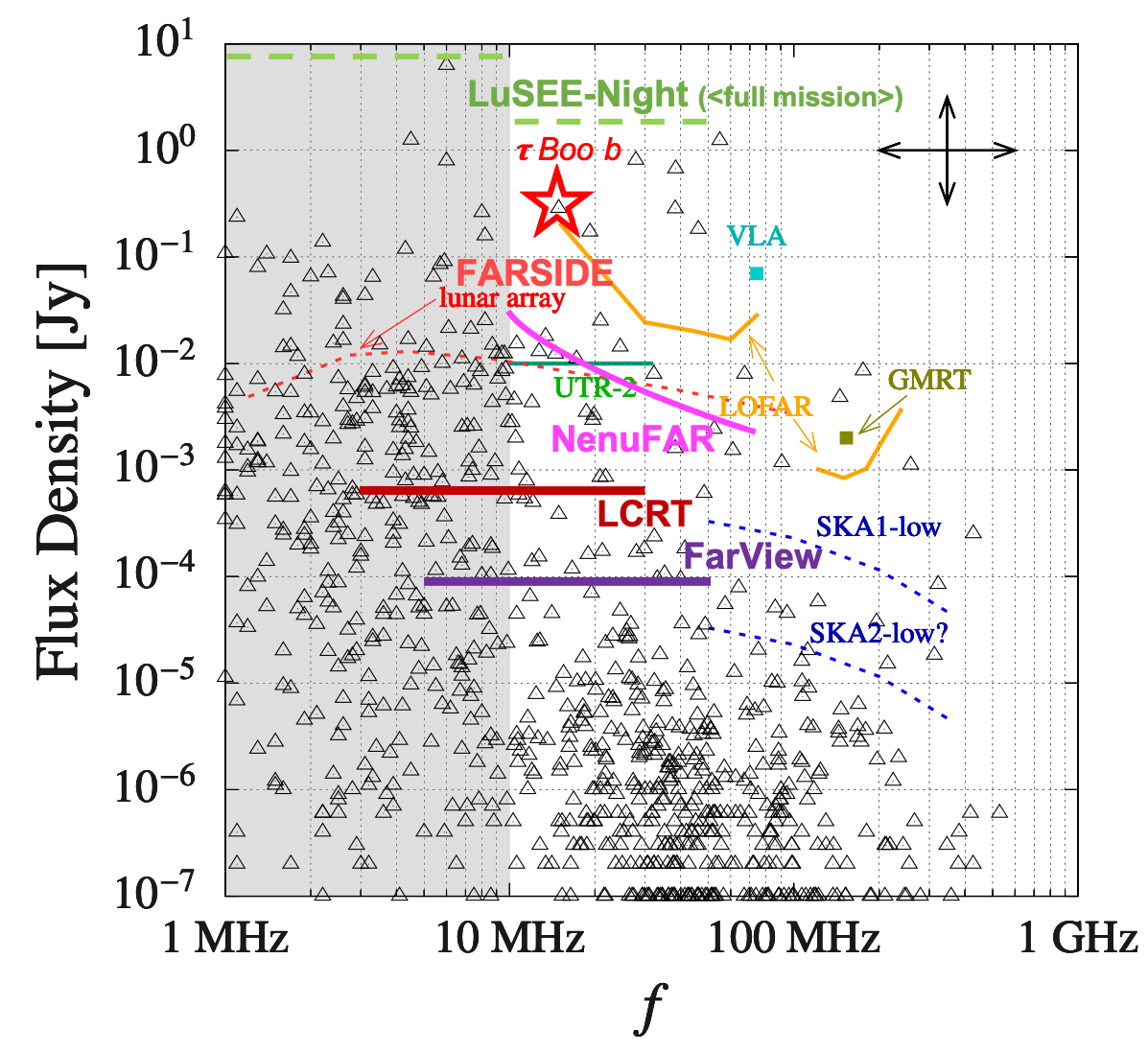}
  \caption{Maximum predicted emission frequency and radio flux density for known exoplanets (triangle symbols; data from [8]). The uncertainties on the predictions are estimated to one order of magnitude for the flux density and a factor of 2-3 for the emission frequency$^{12}$. The sensitivities for the ground-based (\textit{UTR-2}, \textit{LOFAR}, \textit{NenuFAR}, \textit{SKA-Low}, \textit{VLA}, \textit{GMRT}) and space-based (\textit{FARSIDE}, \textit{LCRT}, \textit{FARVIEW}) radio telescopes are labeled (for 10 min of integration time and a bandwidth of 24 MHz). The \textit{FARVIEW} and \textit{LCRT} sensitivity curves are taken from [19] and [2], respectively. The \textit{LuSEE-Night} sensitivity curve is averaged over the entire two-year mission and includes post-processing using a Lomb-Scargle periodogram (see text for details). The possible detection of $\tau$~Boo~b from \textit{LOFAR}$^{23}$ is shown as a red star. This figure is adapted from [9].}
  \vspace{-1.1em}
  \label{fig:predict}
\end{SCfigure}

Starting in the 1970s, a number of ground-based observations have been conducted to find radio emission from gas giant exoplanets, most of which have resulted in clear non-detections. Recent reviews$^{6,7,8,29}$ summarize these observational campaigns. Only gas giant exoplanets (B$_{p}$ $>$ 3.6 G) can be studied from the ground (Figure \ref{fig:Radio_SS}). The low-frequency radio telescopes \textit{LOFAR} and \textit{NenuFAR} are sensitive enough to detect dozens of exoplanets in the radio$^{8,12}$ (Figure \ref{fig:predict}). Motivated by these theoretical predictions, we performed a first of a kind ``Jupiter as an exoplanet'' study using Jupiter radio observations from \textit{LOFAR}$^{21}$ and developed the \texttt{BOREALIS} toolkit to detect these faint extrasolar radio emissions$^{20,21}$. Several large-scale observing campaigns with \textit{LOFAR} and \textit{NenuFAR} are ongoing. Recently, the first possible tentative detection of an exoplanet, the hot Jupiter $\tau$~Boo~b, in the radio was reported using low-frequency beamformed observations from \textit{LOFAR}$^{23}$. The characteristics of the emissions were similar to our ``Jupiter as an exoplanet'' study. The derived field strength, flux density, observed phases, and handedness of the polarization for $\tau$~Boo~b are all in agreement with theoretical predictions$^{1,8,12,18}$. Follow-up observations are already underway to confirm this detection$^{22,24}$. Also, a potential detection of a radio emission burst from the another hot Jupiter system (HD 189733) was detected recently using \textit{NenuFAR} imaging observations$^{30}$. There is great need for more powerful ground- and space-based radio telescopes to complement current ground-based efforts as most exoplanetary systems are too faint to study (Figure \ref{fig:predict}).  

\noindent\textbf{Radio Observations from the Moon:} \\
 \textit{The Solar System as an Exoplanetary System:} The scheduled CLPS radio missions (\textit{LuSEE-Night}, \textit{ROLSES 1 and 2})$^{4}$ and the proposed the \textit{Artemis Tethered Array (ATA)} CLPS mission will be able to study the Sun and the radio-loud Solar System planets (Figure \ref{fig:Radio_SS}) and simulate them as if there were an exoplanetary system. As a proof of concept, we recently measured ``technosignatures'' from Earth with \textit{ROLSES 1}, the first radio telescope on the Moon delivered by a CLPS lander$^{14}$. Therefore, these missions will be pathfinders and allow us to grow our toolkit, similar to our \textit{LOFAR} ``Jupiter as an exoplanet'' work$^{21}$, to study the magnetospheres of many different types of exoplanets (e.g., hot Jupiters to terrestrial planets) with \textit{FarView} and \textit{FARSIDE}. 

Also, \textit{LuSEE-Night} can place meaningful upper limits for exoplanets between 1-10 MHz for the first time (see Figure \ref{fig:predict}). Although individual bursts are too faint for \textit{LuSEE-Night} to detect, several key aspects of the emission will allow for stacking of the observations. Exoplanetary radio emission occurs over large time-scales ($\sim$hours) and bandwidths (10s of MHz), and the emission is periodic (the emission beam will always be pointed towards the observer during the same part of the orbit for tidally locked planets)$^{27}$. This periodicity allows data from hundreds of orbits to be analyzed with a Lomb-Scargle periodogram, a technique recently validated on Jupiter radio observations from \textit{NenuFAR}$^{17}$. This method will obtain an average radio flux upper limit that is comparable to the upper-edge of the theoretical predictions$^{8}$ (Figure \ref{fig:predict}). However, as the flux of exoplanetary radio emissions are expected to be variable due to changes in the stellar wind density, stellar rotation, and the stellar magnetic cycle$^{10,11}$ these limits will be limited in their scope. Nevertheless, these constraints will be extremely beneficial for planning exoplanet observations with future lunar arrays. 

\textit{FarView}$^{19}$ and \textit{FARSIDE}$^{5}$ will revolutionize the study of exoplanetary magnetospheres. To start, \textit{FarView} can study the magnetic fields of a full suite of exoplanetary systems (terrestrial planets to gas giants) since it can probe smaller magnetic field strengths (B$_{p}$ $<$ 3.6 G). Equally importantly, \textit{FarView} can also study an order of magnitude more Jupiter-like planets than current ground-based telescopes (Figure \ref{fig:predict}). This will allow us for the first time to statistically test the dynamo modeling of Jupiter-like planets across a wide range of temperatures (100--2500 K). Based on theoretical models, [25] predict that the magnetic field strengths of close-in planets can be affected by the intense incident stellar flux. Additionally, studying the ground-based radio-loud exoplanets would enable the exploration of the outer parts of the magnetosphere that are not accessible from the ground (Figure \ref{fig:Radio_SS}). Interestingly, the available target-list for \textit{FarView} within the 5-10 MHz frequency range includes a handful of super-Earths and Neptune-like planets (e.g. GJ 1214b). Therefore, \textit{FarView} help refine dynamo modeling for habitable planets in preparation for \textit{FARSIDE}. Mostly importantly, \textit{FARSIDE} can study the magnetic fields of dozens of exoplanets including a handful of the nearest candidate habitable exoplanets$^{5}$ (Figure \ref{fig:predict}). \textit{FARSIDE} will be extremely complementary to the atmospheric studies by \textit{JWST}, \textit{HWO}, and ground-based 30-m class telescopes for understanding the habitability of nearby terrestrial exoplanets. Also, it is worth noting that the \textit{LCRT}$^{2}$ mission concept will have the sensitivity to study exoplanets between 5--30 MHz from the Moon (Figure \ref{fig:predict}). However, only a handful of exoplanets that cross \textit{LCRT}'s beam can be studied since the telescope will have no steering capabilities. In summary, \textit{FarView} and \textit{FARSIDE} will open up a whole new regime in statistical exoplanetary radio science and comparative planetology.

\vspace{0.5em}
{\scriptsize
\noindent \textbf{References:} 
[1] Ashtari et al. 2022, ApJ, 939, 24;
[2] Bandyopadhyay et al. 2021, IEEE, 50100, 1-25;
[3] Brain et al. 2024, RiMG, 90, arXiv:2404.15429;
[4] Burns et al. 2021, PSJ, 2, 44;
[5] Burns et al. 2019, arXiv:1911.08649;
[6] Callingham  et al. 2024, Nat. Astro., 8, 1359;
[7] Grießmeier 2015, ASSL, 411, 213;
[8] Grießmeier 2017, PRE VIII, 285-300
[9] Grießmeier  2018, HAEX, 22;
[10] Grießmeier et al, 2005, A$\&$A, 437, 717;
[11] Grießmeier et al. 2007, P$\&$SS, 55, 618;
[12] Grießmeier et al. 2007, A$\&$A, 475, 359;
[13] Hess et al. 2011, A$\&$A, 531, 29;
[14] Hibbard et al. 2025, arXiv:2503.09842;
[15] Lazio et al. 2019, BAAS, 51, 135;
[16] Lazio 2018, HAEX, 9;
[17] Louis et al. 2025, arXiv:2503.18733;
[18] Mauduit et al. 2023, PRE IX, 103092;
[19] Polidan et al. 2024, AdSpR, 74, 528;
[20] Turner et al. 2017, PRE VIII, 301-313;
[21] Turner et al. 2019, A$\&$A, 624, A40;
[22] Turner et al. 2024, A$\&$A, 688, A66;
[23] Turner et al. 2021, A$\&$A, 645, A59;
[24] Turner et al. 2023, PRE IX, 104048;
[25] Yadav et al. 2017, ApJL, 849, 12;
[26] Zarka 1998, JGR, 103, 20159;
[27] Zarka 2007, P$\&$SS, 55, 598;
[28] Zarka 2018, HAEX, 22;
[29] Zarka et al. 2015, AASKA14, 120;
[30] Zhang et al. 2025, arXiv:2506.07912
}

\end{document}